\documentclass[sigconf]{acmart}

\usepackage{booktabs} 

\begin{document}
\title{Early detection of Crossfire attacks using deep learning}

\author{Saurabh Misra}

\affiliation{%
  \institution{Singapore University of Technology and Design}
}
\email{misra.saurabh1@gmail.com}

\author{Mengxuan Tan}

\affiliation{%
  \institution{Singapore University of Technology and Design}
}
\email{mengxuan_tan2@sutd.edu.sg}

\author{Mostafa Rezazad}

\affiliation{%
  \institution{Singapore University of Technology and Design}
}
\email{mrezazad@gmail.com}

\author{Matthias R. Brust}

\affiliation{%
	\institution{University of Luxembourg}
}
\email{matthias.brust@uni.lu}

\author{Ngai-Man Cheung}

\affiliation{%
  \institution{Singapore University of Technology and Design} 
}
\email{ngaiman_cheung@sutd.edu.sg}


\begin{abstract}
Crossfire attack is a recently proposed threat designed
to disconnect whole geographical areas, such as cities or states, from the Internet. Orchestrated in multiple
phases, the attack uses a massively distributed botnet
to generate low-rate benign traffic aiming to congest selected
network links, so-called target links. The adoption of benign
traffic, while simultaneously targeting multiple network links,
makes the detection of the Crossfire attack a serious challenge.

In this paper, we propose a framework for early detection of Crossfire attack, i.e., detection in the warm-up period of the attack.
We propose to monitor traffic at the potential decoy servers and discuss the advantages comparing with other monitoring approaches.  Since the low-rate attack traffic is very difficult to distinguish from the background traffic, we investigate several deep learning methods to mine the spatiotemporal features for attack detection.    
We investigate Autoencoder, Convolutional Neural Network (CNN) and Long Short-Term Memory (LSTM) Network to detect the Crossfire attack during its warm-up period.
We report encouraging experiment results.


\end{abstract}

%
%

\maketitle

\section{Introduction}

A novel class of an extreme link-flooding DDoS (Distributed Denial of Service) attack \cite{xue2014towards} is the Crossfire attack \cite{kang2013crossfire}.
It is designed to cut off a targeted geographic region from the
Internet by simultaneously targeting a selected set of network
links \cite{gkounis2016interplay}, \cite{gkounis2014towards}. The most intriguing property of this attack is
the usage of legitimate traffic flows to achieve its devastating
impact by making the attack particularly difficult to detect and,
consequently, to mitigate \cite{kang2013crossfire}.

In this paper, we propose a new detection approach that uses  the traffic volume (or intensity) on specific
network regions for any subtle changes on some of the
links. Depending on the resolution of the monitoring scheme,
we show that this leads to an early detection of the attack.
In particular, we argue that monitoring the traffic of the public servers near a target area could facilitate early detection.
We describe several deep learning based methods to extract useful features from the traffic volume: Autoencoder, 
Convolutional Neural Network
(CNN) and Long Short Term Memory (LSTM) network.
We show the feasibility of early detection in our simulation.


\subsection{Crossfire attack}

The Crossfire attack uses a massively large-scale botnet
for attack execution \cite{kang2013crossfire}. The success of the attack depends highly on the network
structure and how the attacker plans and initiates the attack
sequence \cite{zargar2013survey}. The attacker aims to find a set of target links,
which connects to the decoy servers such that if the target links
are flooded, traffic destined to the target area is prevented from
reaching its destination. Reciprocally, access from the target
area to Internet services outside the target area will be cut off.

For the adversary to achieve its goal, it chooses public
servers either inside of the target area or nearby the target area,
which can be easily found due to their availability. The quality
of the attack depends on the specific selection of servers and
the resulting links to be targeted.
\begin{figure}
\includegraphics[height=2in, width=3in]{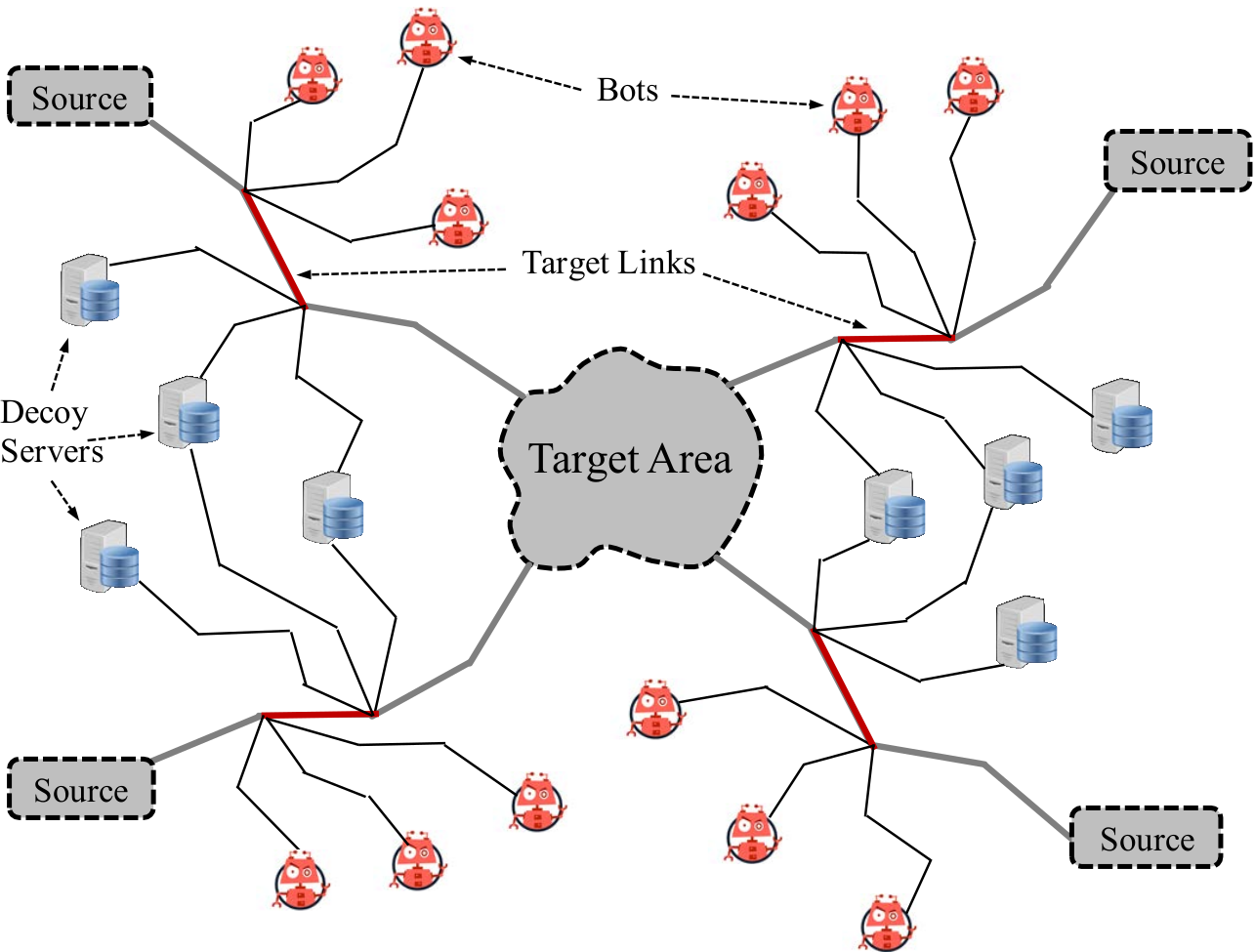}
\caption{The Crossfire attack traffic flows congest a small set of selected network links using benign low-rate flows from bots to publicly accessible servers, while degrading connectivity to the target area.\label{fig:figure1}}

\end{figure}

The Crossfire attack consists of
three phases: (a) the construction of the link map, (b) the
selection of target links, and (c) the coordination of the
botnet. While phases (a) and (b) are sequentially executed only
initially, once triggered phase (c) is executed periodically. Figure 
\ref{fig:figure1} illustrates the dynamics of the Crossfire attack.

\textit{Link map construction}: The attacker creates
a map of the network along the ways from the attacker's
bots to the servers using traceroute \cite{kang2013crossfire}. 

\textit{Target links selection}: After the construction of the link
map, the adversary evaluates the data for more stable and
reliable routes to decide on its selection of the target links.

\textit{Bot coordination}: In the final phase of the attack, the
adversary coordinates the bots to generate low-intensity traffic
and to send it to the corresponding decoy servers. The targeted
aggregation of multiple low-intensity traffic flows on the target
link ideally exhausts its capacity, hence, congesting the link.

Because the Crossfire attack aims to congest the target links
with low-rate benign traffic, neither signature based Intrusion
Detection Systems (IDSs) nor alternative traffic anomaly detectors are 
capable of detecting malicious behavior of individual 
flows. The detectability of the Crossfire attack can be
further hardened by the attacker (a) gradually increasing bot
traffic intensities, (b) estimating the decoy servers' bandwidth
to avoid exceeding their bandwidth, (c) evenly distributing
the traffic over the decoy servers, (d) alternating the set of
bots flooding a target link, and (e) alternating the set of target
links \cite{kang2013crossfire}.
Although these techniques further sophisticate the attack,
the inherent complexities of the attack 
also create substantial execution obstacles, which exposes
the attack to detection vulnerabilities.

\section{Monitoring and Detection Approach}

Considering the described Crossfire attack execution sequence, 
we argue that there are potentially four ways to
detect the attack \cite{mostafa:2018}:
(a) detection at the traffic flows origin, i.e.,
bot sides, (b) detection at the target area, (c) detection at the
target link, and (d) detection at the decoy servers. Following,
we address the advantages and disadvantages each of the four
ways to finally justify our choice for traffic monitoring.
\begin{itemize}
 \item \textit{Detecting at the origin} can be the fastest way to stop an
attack before even it is initiated. However, versatility and
spatial distribution of bots (source of the attack traffic)
makes it the most challenging option.
 \item \textit{Detection at the target area} is the most reasonable
approach as any target areas should be equipped for self-
defense. However, assuming not all decoy servers are
inside the target area, early detection is impossible \cite{xue2014towards}.
 \item \textit{Detection at the target link} might be the simplest form
of detection as simple a threshold based detection system
that could detect the trend of the incoming traffic. However, the locations of the
target links may be unknown to the defender.
 \item \textit{Detection at decoy servers} can be the best approach to
detect Crossfire attack. Assuming the target area is not
far from the decoy servers (3 to 4 hops \cite{kang2013crossfire}) detecting at
the decoy servers might reduce the impact of the attack.
\end{itemize}
Therefore, we propose  to perform detection at the decoy servers,
because it is the exclusive area that the defender can detect
the attack while actively responding to it. To emphasize the
effectiveness of our detection approach at the decoy servers,
we address the question of where is the best location to
probe the network. In a high resolution, this probing can be
placed either at the target link, before or after the target link.
Monitoring a single link as a target link is not considered as
a solution because of two reasons:
\begin{itemize}
 \item Any links can be targeted for an attack. Therefore, there
should be a one-to-one detector for every link in the
network. However, in our proposal, there is only one
detector but many probing points.
 \item Monitoring and detecting based on a single link will fail
in distinguishing between link attack and flash-crowd. 
\end{itemize}
Our main goal is to detect the Crossfire attack without
the need of having the target link information.
Depending on the budget of the adversary, the number of
bots purchased for an attack can be in the range of thousands
to even millions. If the sources of the attack traffic, i.e., bots, are
geographically spread out, the variation of the traffic volume
on most of the links is very small (for many routes, there might
be only one or few attack flow before they are aggregated at
the target link). That leaves only few link closer to a target
link worth to examine. However, the chosen decoy servers
should not be very far away from the target area (if they are
not inside the target area). Since there is a smaller number of
destinations for the attack traffic than the number of sources
of generating them, it can be assumed that the variation of the
volume of the traffic caused by the attack traffic on the links
after the target link is higher than the links before the target
link (even at the edge of the network). Therefore, we propose to 
monitor links around servers or data centers that can result in
more successful detection than around clients.

The approach of evenly distributing the traffic for decoy
servers \cite{kang2013crossfire}, might even support the above reasoning and rather
make it simpler to detect some variation in the traffic volume
on several links. The important element in this method is to
be able to monitor the traffic at several links and send the
information to a detector for decision making.

\section{Attack Characteristics and Experiment Set-up}
Since our focus is on the detection of the
attack, we ignore the first few steps of the Crossfire attack
such as link map construction, finding link persistence, or target
link selection. We assume that all attack preparations have
been made and the attacker is ready to attack.

When the preparations have been made, to bring down the target link, the bot-master initiates the attack by sending the attack order to the Command and Control (C\&C) server or some selected peers depending on the structure of the botnet. Bots usually update
each other in a polling or pushing mechanism.

When designing Crossfire detection mechanisms, an often
ignored part of the Crossfire attack is the phase from the attack
initiation and the successful impact of the attack \cite{kang2013crossfire}. This often
ignored part of the Crossfire attack, which we call it warm-up
period, is the time difference between the time of the first bot-
flow of the attack reaches the target link and the moment the
target link is down. By definition, the attack actually happens
at the end of the warm-up period when the target links are
down. Since, reaching a zero time warm-up period is hard,
this period can be used for early detection and before the
attack successfully takes place.

In fact, for several reasons reaching a zero warm-up time is
hard. One reason is the dynamic delay of packet arrival
at the target link. That could be because of variations of hop
distances from bots to target link, or the delay in receiving
attack order from the adversary. Any sudden significant change
on traffic volume can be detected by firewalls and IDSs.
Therefore, adversaries gradually increase the attack traffic
volume to prevent being detected.

Another important parameter for generating the bot traffic is the
duration of the attack. The attack duration is the time difference between the end of the warm-up period and the end of the attack. Usually, bot-masters (adversaries) tend
to reduce the duration of the attack to prevent being detected.
In the case of the Crossfire attack, a rolling mechanism
is introduced to keep the attack at the data plane (evade
activating control plane which redirects the traffic) \cite{kang2013crossfire}. In
the rolling scheme, a set of target links are only used for a
specific period of time before switching to another set of
target links. The duration they used in the rolling scheme is
3 minutes or less where 3 minutes is the keep-alive messages time interval
for the BGP algorithm. 


To generate data for the training Crossfire detectors, we simulate link utilizations of a network with 80 decoy servers under attack in two different conditions. In the first case all the 80 decoy servers are under attack and in the other case 70 decoy servers which are randomly selected each time are under attack. We also study the case when not all decoy servers are under attack as we cannot know in advance exactly which all servers would the attacker target. We sample the average link utilization every 1 minute. When not under an attack, the servers have a background traffic which is simulated by a random Gaussian process assuming the server is communicating with many clients. The background traffic for each server has a different mean between 100Kbps-150Kbps and a standard deviation between 0.45Kbps-2.45Kbps.

\begin{figure}
\includegraphics[height=2in, width=3.5in]{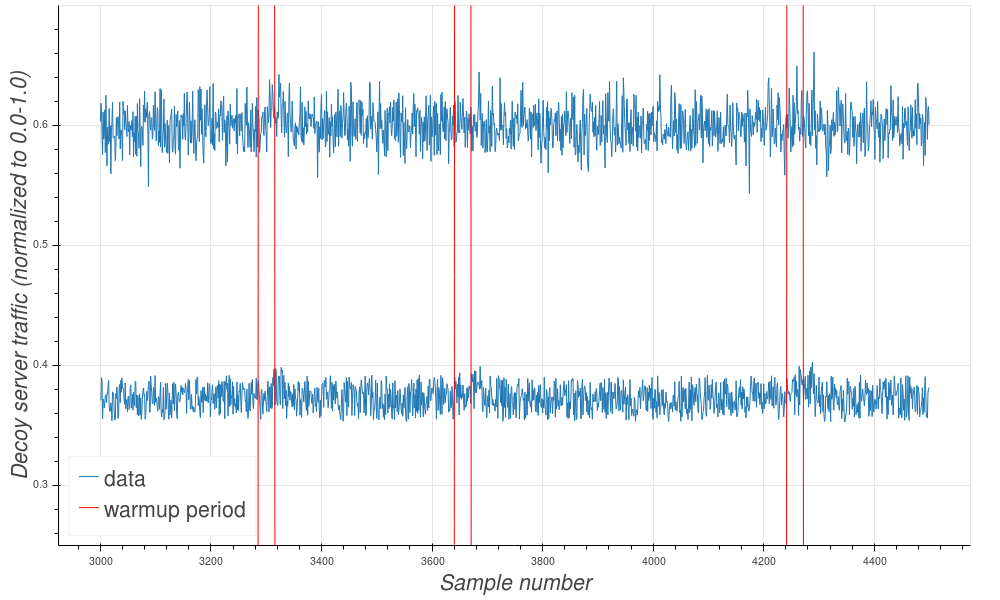}
\caption{The simulated low-rate network utilization data for 2 decoy server locations when all servers are under attack.
The link utilization is normalized to a range of 0-1 with 0 and 1 corresponding to the minimum and maximum of all measurements respectively.
Note that 
the low rate traffic in the warm up period is very difficult to distinguish from the background traffic.
\label{fig:figure2}}

\end{figure}

To simulate an Crossfire attack, an extra attack traffic is added to the background traffic. The attack starts with a warm-up period where the attack traffic slowly increases in intensity. The warm-up duration is 30 minutes (samples). The warm-up period corresponds to a randomized ramp function. When the attack manages to flood the target link, the attack traffic reaches its peak and stays there till the attack is over. The intensity of the low rate network traffic from the bot for the attack ranges from 0.43Kbps-2.2Kbps.
The dataset for both the conditions has 6000 separate attack instances. The data is labeled true only for the warmup period as we attempt to do early detection before the attack has flooded the target links.
Figure \ref{fig:figure2} illustrates the link utilization for 2 decoy servers during 3 separate attack instances when all the decoy servers are under attack.
Note that it is difficult to distinguish the attack traffic from the background traffic.

%
%

\section{Crossfire detection with deep learning}

Crossfire attack poses numerous challenges for security
researchers and analysts both in detection and mitigation as
the packets streaming from bots in the network are seemingly
legitimate. While the objective of the Crossfire attack is to
deplete the bandwidth of specific network links, a distinct
traffic flow between each bot to server, i.e., "bot-to-server"
is usually very less intensive flow, and consumes a limited
bandwidth at each link. Thus detecting a single flow (or very
few number of them) at a link is hard to detect and filter. On
the defender's side, Traffic Engineering (TE) is the network
process that reacts to link-flooding events, regardless of their
cause \cite{liaskos2016novel}. The goal of an attacker is to hide the variation of
traffic bandwidth as much as possible from the TE module.

Following this direction, we leverage  
 deep learning approaches to detect the Crossfire attack
from available traffic data. 
Recently, 
deep learning has achieved
breakthrough results
in many natural language processing and computer vision problems \cite{Krizhevsky:2012,He:2016,Toan:2016}.  Here we investigate its usefulness and  performance for traffic data analysis to detect Crossfire attack in the early stage.


\subsection{Deep Autoencoders}







\begin{figure}
\includegraphics[height=1.3in, width=1.8in]{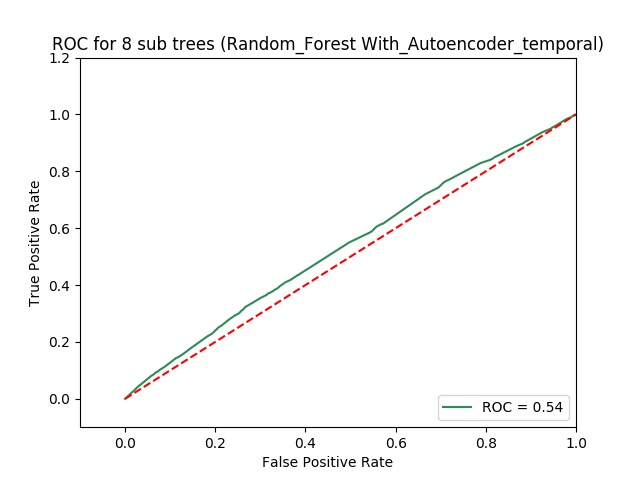}
\caption{Performance of Random Forest on temporal data extracted by deep-autoencoder.\label{fig:temp_fig}}
\end{figure}

We investigate using 
deep-autoencoder \cite{hinton2006reducing} to extract intrinsic low-dimensional information from the traffic. 
This is followed by Random Forest 
for anomaly detection.

Here we propose a deep-autoencoder structure to exploit spatiotemporal information.
We uses a window of 5 consecutive time steps by concatenating data from 5 consecutive samples.
Each sample contains 80 measurements that
correspond to the traffic volumes of the 80 decoy servers in the simulated network.
Thus the input dimension of the data for the autoencoder is  400.
The autoencoder contains 2 hidden layers and its structure is as such: (400)-390-370-390-400.
Figure \ref{fig:temp_fig} shows the area under the Receiver Operating Characteristic (ROC) curve of this approach.

\subsection{Convolutional Neural Network}
A Convolutional Neural Network (CNN) can be utilized for detecting by considering the data as a 2 dimensional grid spanning across time and decoy server dimensions. The dataset is divided into separate windows with 80 servers we are monitoring as columns and across 15 time steps as rows. We label the entire window as an attack window if the warmup period of the attack occurs in at least 5 out of 15 time steps. 
\begin{figure}
\includegraphics[width=3.5in]{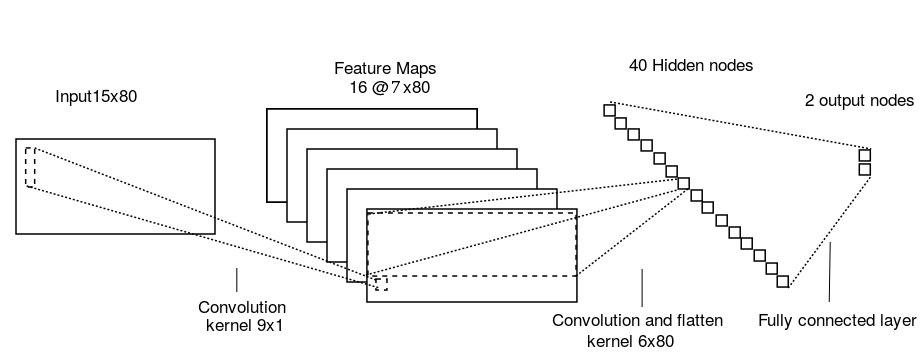}
\caption{The CNN architecture used to detect crossfire attacks. The output layer is a binary predictor predicting the two categories of attack and non-attack\label{fig:cnn}}

\end{figure}
The CNN is arranged in two separate convolutional operations. The first step learns a convolutional filter that spans only across the time axis(rows) and has dimension of 1 across the server axis(columns). This filter is expected to learn the pattern for the attack only in the time axis independent of the servers. We call this as a temporal filter.  The second convolutional filter spans across all the 80 servers and a few rows. We call this as the spatial filter as it extends through space for all the servers. This filter is expected to discover the correlation between different servers as they are under attack at the same time. The intuitive reasoning behind such a network configuration was validated by a hyperparameter search across various configurations. The activation function used is ReLu and to improve network performance each convolutional step is batch normalized. A final fully connected layer with softmax activation does a binary classification of the window as attack or non-attack. The network architecture is demonstrated in figure ~\ref{fig:cnn}.

The shape of the temporal filter is 9x1x16 (height x width x depth). The spatial filter dimension is 6x80x20 (height x width x depth). The filter strides for each case is 1. An Adam optimizer with learning rate of 0.3x10\textsuperscript{-5} is used to train till the early stopping condition. The dataset is divided into Training, Validation and Test set in the ratio of 70:20:10.

The results on the test dataset for two different attack conditions is shown in Table ~\ref{tab:table2}.
\begin{table}[h!]
  \begin{center}
    \caption{Performance of CNN.}
    \label{tab:table2}
    \begin{tabular}{l|c|c|r} 
       $\textbf{Servers under attack}$ & $\textbf{Precision}$ & $\textbf{Recall}$ & $\textbf{F1 Score}$ \\
      \hline
      80/80 & 0.74 & 0.97 & 0.84\\
      \hline
      70/80 & 0.759 & 0.788 & 0.773
    \end{tabular}
  \end{center}
\end{table}
\subsection{Long Short-Term Memory Network}

\begin{figure}
\includegraphics[width=3.5in]{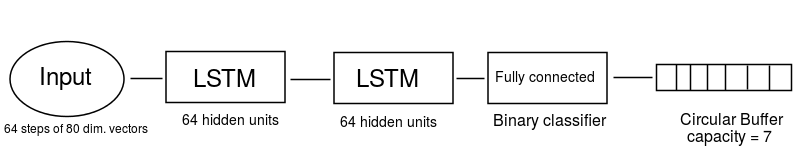}
\caption{The LSTM architecture for detecting Crossfire attacks\label{fig:lstm}}
\end{figure}

An LSTM\cite{Hochreiter:1997:LSM:1246443.1246450} is used as a sequence classifier for detecting whether each time sample is a warm-up period of Crossfire attack or not. We use a stacked LSTM configuration having 2 consecutive LSTM cells as shown in Figure~\ref{fig:lstm}. Each of the LSTM cells have 64 hidden units. Each output sample from stacked LSTM is classified into an attack or non-attack by a fully connected layer. The input to the LSTM network are windows of 64 time samples of 80 dimensional vectors for decoy traffic. Adam optimizer is used to train the network with learning rate as 0.001 till the Early stopping condition.

To improve on the per sample prediction of learned LSTM and also to exploit the fact that warm-up samples are consecutive, we keep the history of the per sample prediction of LSTM Network in a circular buffer of length 7. The scalar value in the buffer cell is 1 if there is an attack and 0 otherwise. Only if all the samples in the buffer are attacks, do we finally predict an attack. This increases latency in predicting an attack but dramatically increases attack detection accuracy. With a buffer of length 7, we find that we can predict almost perfectly for the warm-up period with a maximum latency of 13 samples. The precision of the detector is 1.0 and recall is 0.998, shown in Table ~\ref{tab:table4}. We can also trade-off between the latency of detection and the performance of detection by changing buffer size.
\begin{table}[h!]
  \begin{center}
    \caption{Performance of LSTM.}
    \label{tab:table4}
    \begin{tabular}{l|c|c|r} 
      $\textbf{Servers under attack}$ & $\textbf{Precision}$ & $\textbf{Recall}$ & $\textbf{F1 Score}$ \\
      \hline
      80/80 & 1.00 & 0.998 & 0.999\\
      \hline
      70/80 & 0.995 & 0.964 & 0.979\\
    \end{tabular}
  \end{center}
\end{table}
\section{Conclusions}
The Crossfire attack is considered to be one of the most
difficult target-area link-flooding attacks to be detected. The
attack uses a distributed botnet to generate multiple low-rate benign traffic flows aiming to congest
selected network link with the ultimate goal to disconnect the
target area from the Internet.
In this paper, 
we have demonstrated the effectiveness of monitoring link utilizations at the decoy servers  to detect Crossfire attacks. Out of the different deep learning methods we have investigated, we find that the LSTM detection approach 
achieves encouraging performance under different attack conditions.  For future work, we investigate the performance under different network structures.

\bibliographystyle{ACM-Reference-Format}
\bibliography{sample-bibliography} 

\end{document}